\documentclass[aps,prd,twocolumn,showpacs,superscriptaddress,footinbib]{revtex4-2} 
\usepackage{amssymb}
\usepackage{amsmath}
\usepackage[colorlinks=true,linkcolor=blue,citecolor=blue, urlcolor=blue]{hyperref} 
\usepackage{etoolbox} 
\makeatletter
\patchcmd{\subsubsection}{\itshape}{\bfseries}{}{}
\makeatother
\usepackage{tikz}
\usepackage{graphicx}
\usepackage[utf8]{inputenc}
\definecolor{uibred}{RGB}{167, 38, 47}

\begin{document}

\title{Coupling renormalization flow 
in the strongly interacting regime of an\\  asymptotically free quantum field theory in four dimensions}

\author{J{\"u}rgen Berges}

\author{Razvan Gurau} 

\author{Hannes Keppler} 

\author{Thimo Preis} \email{preis@thphys.uni-heidelberg.de}

\affiliation{Institut f\"{u}r Theoretische Physik, Universit\"{a}t Heidelberg, 69120 Heidelberg, Germany}

\begin{abstract}
We consider a scalar quantum field theory with global $O(N)^3$ symmetry in four Euclidean dimensions and solve it numerically
in closed form in the 
large-$N$ limit. For imaginary tetrahedral coupling the theory is asymptotically free, with stable and real quantum effective action. We demonstrate the dynamical build-up of a strong interaction as the correlation length increases in a regime where the coupling renormalization flow  remains well defined in the infrared. 
This is in contrast to perturbative results of asymptotically free theories, which predict that the coupling becomes ill defined in the infrared, like in quantum chromodynamics. These properties make the model an important laboratory for the study of strong-coupling phenomena in quantum field theory from first principles. 
\end{abstract}
 
\maketitle

\section{Introduction and overview}
Asymptotically free quantum field theories are a cornerstone in the fundamental description of nature. A prominent example is the theory of quantum chromodynamics (QCD) in the Standard Model of particle physics~\cite{Gross:1973id,Politzer:1973fx}. While the high-momentum (ultraviolet) behavior of the theory is perturbatively accessible, the scale-dependent (``running'') coupling increases toward low momenta and becomes ill defined in the infrared. The divergence of the coupling at a finite infrared momentum predicted by perturbation theory illustrates the dynamical generation of a nonperturbative scale by quantum fluctuations. 
Such behavior is characteristic for the perturbative analysis of asymptotically free theories, and it would be highly valuable to establish a nonperturbative example where the coupling is well defined and can be followed all the way from the weakly coupled high-momentum regime to the strongly interacting infrared.

\begin{figure}[t!]
\includegraphics[width=0.99\columnwidth]{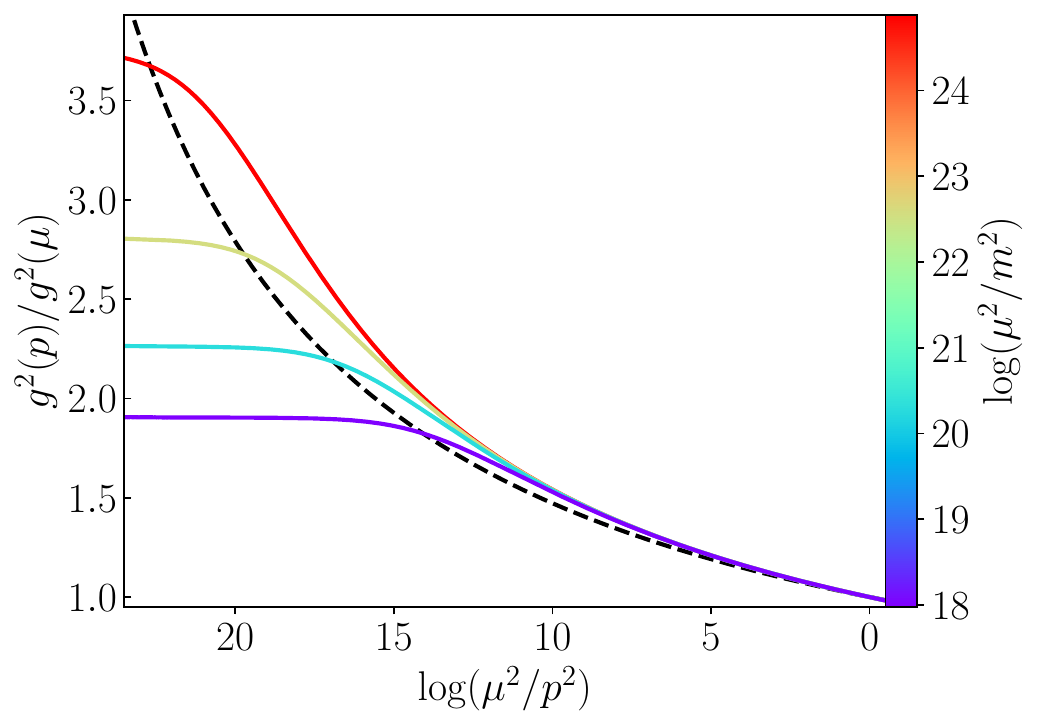}\caption{Flow of the squared tetrahedral coupling $g^2$ with momentum $p$ for various renormalized masses $m^2$ (color scale), with $g(\mu)=20$ at the renormalization scale $\mu$. The two-loop perturbative running is mass independent and represented as dashed.}
\label{fig.Zflow}
\end{figure}

In this work we investigate the large-$N$ limit of a four dimensional scalar quantum field theory with global $O(N)^3$ symmetry introduced in Ref.~\cite{Berges:2023rqa}. The model has three independent quartic couplings, whose perturbative renormalization flow, which encodes how 
the physical couplings change with the momentum scale due to quantum corrections, has been analyzed in Refs.~\cite{Berges:2023rqa,Giombi:2017dtl,Benedetti:2019eyl}. In four Euclidean dimensions the couplings exhibit asymptotic freedom in a regime governed by the flow of an imaginary tetrahedral coupling $i g(p)$~\cite{Berges:2023rqa}. In turn, the tetrahedral coupling diverges in perturbation theory at a finite infrared momentum scale $\mu^*_{\mathrm{pert}}$. A corresponding perturbative behavior is found, in particular, also for the running coupling in quantum chromodynamics~\cite{Gross:1973id,Politzer:1973fx}. The perturbative scale-dependence of the squared coupling of our model is represented by the dashed curve in Fig.~\ref{fig.Zflow}. 

This is to be contrasted with the nonperturbative \mbox{large-$N$} renormalization flow of the coupling we obtain in this work, which is displayed by the solid (colored) curves for various values of the renormalized mass in Fig.~\ref{fig.Zflow}. Contrary to the perturbative prediction, the full coupling is found to depend on the renormalized mass or, equivalently, the inverse correlation length. 
Our results show the generation of a strong interaction by quantum fluctuations, which builds up as the correlation length is increased. 
The running coupling remains well defined and finite in the infrared for renormalized masses above a threshold value $m^*$. Approaching this threshold from above, the growth behavior of the running coupling defines a strong-interaction scale $\mu^*$, which plays the role of a nonperturbative generalization of $\mu^*_{\mathrm{pert}}$. We find  $\mu^*$ to be larger than the scale determined from two-loop perturbation theory. While masses above $m^*$ allow us to investigate the running coupling in a controlled way for momenta even below $\mu^*$, we emphasize that the two scales should not be identified, and we find $m^*$ to be significantly smaller than $\mu^*$. Disentangling these scales allows one to follow the coupling flow from an ultraviolet Gaussian fixed point to a strongly interacting but well defined infrared in quantum field theory, opening up a new pathway for the investigation of strong-coupling phenomena in four dimensions from first principles.

\section{$\boldsymbol{O(N)^3}$ symmetric tensor field theory} 
Following Ref.~\cite{Berges:2023rqa}, we consider a scalar (under rotations) field $\bar{\varphi}_{\mathbf{a}=(a^1, a^2 ,a^3)}$ with $a^{i=1,2,3}=1,\dots,N$ transforming as a real 3-index tensor in the trifundamental representation of $O(N)^3$~\cite{Carrozza:2015adg, Klebanov:2016xxf, Giombi:2017dtl}. 
The model is defined by the bare (classical) action
\begin{align}
    S[\bar\varphi] =  \int& d^4 x \bigg\{ \frac{1}{2} \bar\varphi_{\mathbf{a}}(x)(-\partial^2 +\bar{m}^2) \bar\varphi_{\mathbf{a}}(x) \nonumber\\
    &+\frac{1}{4}\left(\bar{g}_1 \hat{P}^{(1)}_{\mathbf{a} \mathbf{b};\mathbf{c} \mathbf{d} } + \bar{g}_2  \hat{P}^{(2)}_{\mathbf{a} \mathbf{b};\mathbf{c} \mathbf{d} }+i\bar{g} \hat{\delta}^t_{\mathbf{a} \mathbf{b}\mathbf{c} \mathbf{d} }\right)\nonumber\\
    &\hphantom{+}\times \;\bar\varphi_{\mathbf{a}}(x) \bar\varphi_{\mathbf{b}}(x) \bar\varphi_{\mathbf{c}}(x) \bar\varphi_{\mathbf{d}}(x) \bigg\}\;
    \label{eq:S}
\end{align}
in four Euclidean dimensions. 
Here $\bar{m}$ is the bare mass parameter, and we take the bare quartic couplings $\bar{g}_1$, $\bar{g}_2$ and $\bar{g}$ to be real such that $i \bar{g}$ in Eq.~(\ref{eq:S}) is purely imaginary. The three interaction terms in the action stem from the three $O(N)^3$ invariant contraction patterns (``pillow'', ``double-trace'' and ``tetrahedral'')
\begin{align}
    \hat{\delta}^p_{\mathbf{a} \mathbf{b}; \mathbf{c}\mathbf{d}}&= \frac{1}{3N^2} \sum_{i=1}^3 \delta_{a^i c^i} \delta_{b^i d^i} \prod_{j\neq i} \delta_{a^jb^j} \delta_{c^jd^j} \; ,\nonumber \\
    \hat{\delta}^d_{\mathbf{a}\mathbf{b};\mathbf{c} \mathbf{d}} &=N^{-3} \prod_{i=1}^3 \delta_{a^ib^i} \prod_{j=1}^3 \delta_{c^jd^j} \; ,  \\
    \hat{\delta}^t_{\mathbf{a} \mathbf{b} \mathbf{c} \mathbf{d}} &= N^{-3/2} \delta_{a^1 b^1} \delta_{c^1 d^1} \delta_{a^2 c^2} \delta_{b^2d^2} \delta_{a^3 d^3} \delta_{b^3 c^3} \; , \nonumber
\end{align}
which relate to the orthonormal projectors  $\hat{P}^{(1)}=3(\hat{\delta}^p-\hat{\delta}^d)$ and $\hat{P}^{(2)}=\hat{\delta}^d$ appearing in Eq.~(\ref{eq:S}).

\section{Renormalized correlation functions}
Physical observables can be obtained from the renormalized large-$N$ quantum field theory~\cite{Blaizot:2003an,Berges:2005hc,Berges:2004hn,Aarts:2004sd,Reinosa:2006cm,Fejos:2007ec,Blaizot:2010zx,Blaizot:2021ikl}. 
We aim to compute correlation functions representing expectation values of products of quantum fields, specifically the renormalized two-point correlation function or full propagator $G_{\mathbf{a}\mathbf{b}}(x,y)$. In the tensor field theory this computation can be achieved in closed form in the large-$N$ limit. By contrast, in asymptotically free theories like quantum chromodynamics a resummation of the large-$N$ planar Feynman diagrams~\cite{tHooft:1973alw,Brezin:1977sv} is out of reach. This gives us unique access also to the nonperturbative infrared behavior of our large-$N$ theory. 

The renormalized field correlation functions are obtained after imposing renormalization conditions. Two of them concern the full propagator and we write
\begin{equation}
\begin{gathered}
    G^{-1}(0) = m^2\;,\quad \frac{G^{-1}(\mu)-G^{-1}(0)}{\mu^2} =1 \, .
\end{gathered}
\label{eq.renormCond_prop}
\end{equation}
The first condition fixes the renormalized mass $m$ at zero momentum. The second one specifies the wave function renormalization
\begin{equation}
\label{eq.Zofp}
Z(p) = \frac{G^{-1}(p)-G^{-1}(0)}{p^2} \, , 
\end{equation}
at some high momentum scale $\mu$ (the renormalization scale)
to $Z(\mu) = 1$. Three additional renormalization conditions fix the three couplings at the same renormalization scale $\mu$; in particular, the tetrahedral coupling is fixed to a given $g(\mu)$, and the results for the running coupling are presented as a ratio as in Fig.~\ref{fig.Zflow}.
 In the large-$N$ limit the running of the renormalized tetrahedral coupling is entirely encoded in the scale dependence of the wave function renormalization (\ref{eq.Zofp}) as (see also the Appendix)~\cite{Benedetti:2019eyl}
\begin{equation}
\label{eq.renormGamma4}
 g(p) = \frac{g(\mu)}{Z^2(p)} \;.
\end{equation}

With these renormalization conditions
the full \mbox{large-$N$} propagator in the $O(N)^3$-symmetric regime, where $G_{\mathbf{a}\mathbf{b}}= G \delta_{\mathbf{a}\mathbf{b}}$, is determined in momentum space by 
\begin{align}
    G^{-1}(p) & = p^2+ m^2+ g^2(\mu) \int \frac{d^4 q}{(2\pi)^4}\frac{d^4 k}{(2\pi)^4}  G(q) G(k) \,  \nonumber\\ 
 &   \times \bigg[ G(p+q+k) - G(q+k) \nonumber\\
 & \quad - \frac{p^2}{\mu^2} \Big( G(\mu+q+k) - G(q+k) \Big) \bigg]. 
 \label{eq.sdeq_LO}
\end{align}
The self-consistent solution of this equation may be viewed as resumming infinitely many perturbative contributions in the quartic couplings. It is remarkable that Eq.~(\ref{eq.sdeq_LO}) contains in closed form all the relevant phenomena in the asymptotically free regime and in the strongly interacting infrared we are addressing. The solution for the nonperturbative propagator $G(p)$ determines the wave function renormalization $Z(p)$ and the coupling $g(p)$ according to Eqs.~(\ref{eq.Zofp}) and (\ref{eq.renormGamma4}). The behavior of the other couplings $g_1$ and $g_2$ does not enter the solution of Eq.~(\ref{eq.sdeq_LO}). Their running is in turn dictated by the momentum dependence of $g(p)$ and has been shown to be perturbatively already well defined in the infrared in Ref.~\cite{Berges:2023rqa}. Even at next-to-leading order in the large-$N$ expansion one only encounters further tadpole corrections which are compensated by the mass renormalization~\cite{Berges:2023rqa}.

We iteratively solve Eq.~(\ref{eq.sdeq_LO}) until convergence is observed (see details in the Appendix). We verified that the relevant physical results are insensitive to changes in the momentum discretization for the numerical parameters explored in this work.

\section{Perturbation theory} 
We first summarize the two-loop perturbative results as detailed in the Appendix, which predict [see Eq.~\eqref{eq.Zpert}]
\begin{equation}
\begin{split}\label{eq:pertflow}
g_{\mathrm{pert}}^2(p) & = \frac{g^2(\mu)}{1 +  \frac{2 g^2(\mu)}{(4\pi)^4} \log\left(\frac{p^2}{\mu^2}\right)}  \; .
\end{split}
\end{equation}
This perturbative result exhibits a pole at the finite infrared momentum scale 
\begin{equation}
\mu^*_{\mathrm{pert}} = \mu \, e^{-(4\pi)^4/(4 g^2(\mu))} \, .
\label{eq:mustar}
\end{equation}
This behavior may be contrasted with the perturbative coupling flow of a familiar scalar quantum field theory with single-component field $\phi$ and quartic interaction term $\lambda \phi^4$ in four dimensions. In that case the renormalization flow is described by the beta-function \mbox{$\beta_{\lambda}\sim \lambda^2$}, leading to the scale-dependent coupling
$\lambda(p)=\lambda(\mu)\left(1-K \lambda(\mu)\log(p^2/\mu^2)\right)^{-1}$ for some constant $K>0$. Comparing to Eq.~(\ref{eq:pertflow}), one observes that a sign flip in the denominator transforms the UV Landau pole of the $\phi^4$ model into an IR pole of an asymptotically free theory -- a well known feature of the $\phi^4$-model with negative coupling, $\lambda <0$~\cite{Symanzik:1971vw,Symanzik:1973hx}. However, such a model with repulsive interaction has classically an unbounded spectrum and is therefore considered unstable. Recently this conclusion has been reinvestigated in the context of $\mathcal{PT}$-symmetry~\cite{Bender:1998ke,Lawrence:2022afv,Romatschke:2023fax}. In contrast, our theory is bounded from below due to the two additional (positive semidefinite) quartic couplings and, importantly, the two-loop beta-function $\beta_{g,\mathrm{real}}\sim g^3_{\mathrm{real}}$ changes sign for an imaginary coupling $g_{\mathrm{real}} \to ig $, not a negative one. The beta-function of our tensor field theory starts at cubic order in the tetrahedral coupling because its flow is driven solely by the wave function renormalization.

\begin{figure}[t!]
\includegraphics[height=0.7\columnwidth]{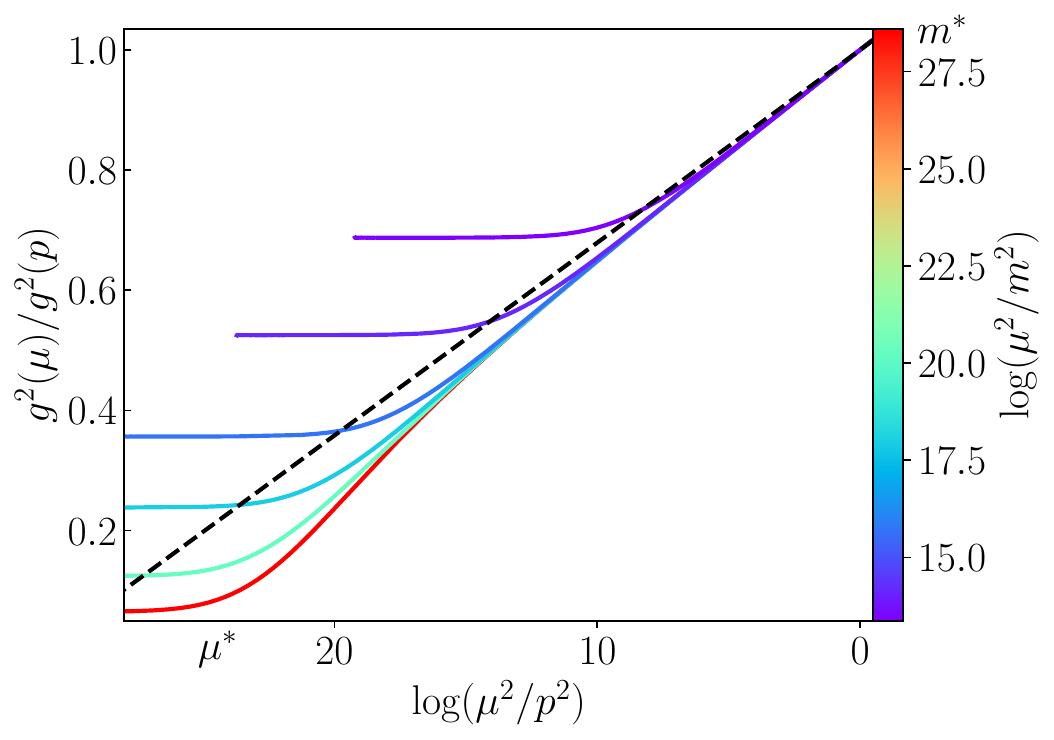}\caption{Inverse of the squared tetrahedral coupling for different renormalized masses (color scale). The perturbative result is displayed as a black dashed line.
}
\label{fig.g2Invers}
\end{figure}

\section{Nonperturbative flow of the tetrahedral coupling} The running of the tetrahedral coupling is controlled by the wave function renormalization and, gathering Eqs.~\eqref{eq.Zofp} and~\eqref{eq.renormGamma4}, we obtain
\begin{equation}\label{eq.gofp}
    \frac{g^2(\mu)}{g^2(p)} = Z(p)^{4} = \left[\frac{G^{-1}(p)-G^{-1}(0)}{p^2} \right]^4 \;.
\end{equation}
The full large-$N$ result for the scale-dependent coupling is given for a wide range of renormalized masses (see color code) in Fig.~\ref{fig.Zflow} and Fig.~\ref{fig.g2Invers}. In Fig.~\ref{fig.g2Invers}, we plot its inverse on a logarithmic momentum scale for a broader range of masses to illustrate the deviation from the perturbative result (\ref{eq:pertflow}), which is represented by a strictly straight dashed line. At large momenta, the perturbative and nonperturbative solutions agree increasingly well, consistent with the prediction of asymptotic freedom. However, the slope of the nonperturbative flow becomes steeper toward the IR such that the coupling grows faster than the perturbative one at intermediate scales. 

\begin{figure}[t!]
\vspace{0.09cm}
\includegraphics[height=0.6950\columnwidth]{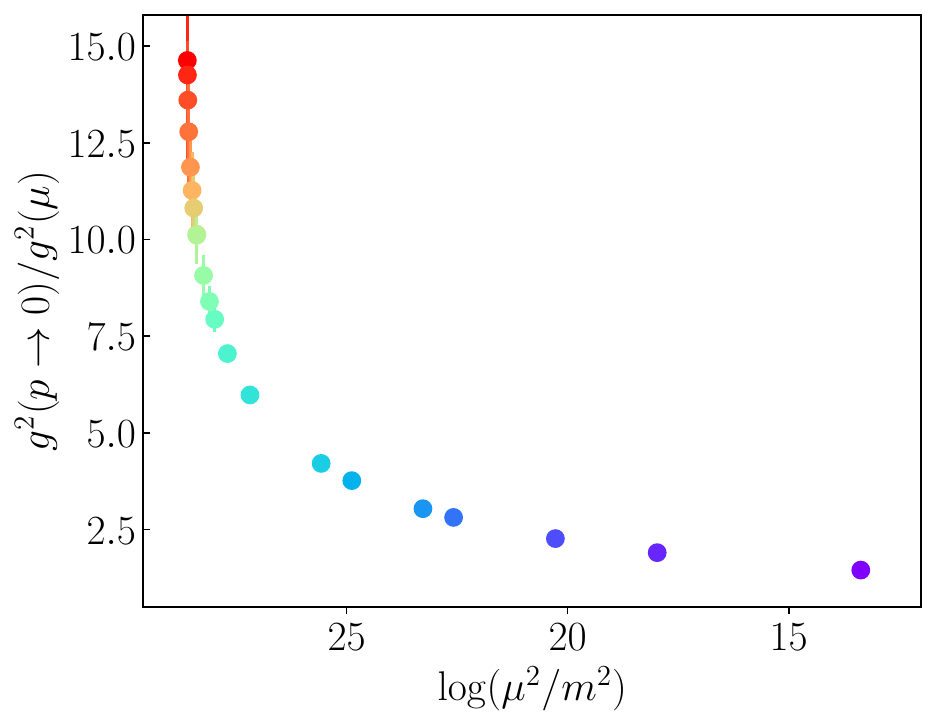}\caption{Limiting value of the squared tetrahedral coupling in the infrared for varying mass.
}
\label{fig.gIR_vs_m}
\end{figure} 

While the perturbative result is insensitive to the renormalized mass, the full coupling is seen to depend on it. The coupling reaches a mass-dependent finite value in the deep infrared for masses above a threshold value. As we decrease the renormalized mass, the limiting value of the tetrahedral coupling grows, as depicted in Fig.~\ref{fig.gIR_vs_m}~\footnote{The numerical error is estimated by using once $G^{-1}(0)=m^2$ and once $G^{-1}(\Lambda_{\mathrm{IR}})$, with $\Lambda_{\mathrm{IR}}$ the numerical IR cutoff, to subtract the momentum independent part in Eq.~\eqref{eq.gofp}.}.
The observed behavior suggests that, for any given $g(\mu)$, there exist a finite mass $m^*$ for which the tetrahedral coupling diverges at a finite IR momentum scale $\mu^*$.
We stress that the mass scale $m^*$ should not be identified with the dynamically generated scale $\mu^*$. We can estimate $m^*$ from Fig.~\ref{fig.gIR_vs_m} and the momentum scale $\mu^*$ by extrapolating the envelope of the curves in Fig.~\ref{fig.g2Invers}. We find that the dynamically generated scale $\mu^*$ is significantly larger than the mass scale $m^*$, which in turn is much larger than the perturbative scale $\mu^*_{\mathrm{pert}}$~\footnote{For $g(\mu)=20$, as employed for the figures, we find $\log(\mu^2/(\mu^*)^2)\simeq 25.2$,
$\log(\mu^2/(m^*)^2)\simeq 28.6$ and $\log(\mu^2/(\mu^*_{\mathrm{pert}})^2)\simeq 31.2$.}. These values of $m^*$ and $\mu^*$ are highlighted in Fig.~\ref{fig.g2Invers}. For completeness, the corresponding results for the propagator are discussed in the Appendix.

\section{Relation to other tensor field theories} In order to put our results into context, we compare to similar tensor models in the literature. First, one can consider the $O(N)^3$ model in $4-\epsilon$ dimensions~\cite{Giombi:2017dtl}. This model has a nontrivial (so-called ``melonic") fixed point with couplings of order~$\sqrt{\epsilon}$. However, while the tetrahedral coupling is real, the pillow and double trace couplings are imaginary at the fixed point, and the resulting conformal field theory is unstable as it has a primary operator in the principal series~\cite{Kim:2019upg,Benedetti:2021qyk}. Alternatively, one can consider a long-range version of the $O(N)^3$ model in $d<4$ dimensions \cite{Benedetti:2019eyl,Benedetti:2020sye}. Picking the marginal scaling for the propagator and an imaginary tetrahedral coupling, one obtains a line of infrared fixed points (indexed by the exactly marginal tetrahedral coupling), which are stable and correspond to well defined (and presumably unitary~\cite{Benedetti:2019ikb}) large-$N$ conformal field theories.
More generally, the renormalization group fixed points for tensor field theories give rise to a new family of conformal ``melonic'' field theories which can be studied analytically~\cite{Benedetti:2018goh,Benedetti:2019rja, Benedetti:2021wzt,Prakash:2017hwq, Benedetti:2017fmp, Giombi:2018qgp, Witten:2016iux, Gurau:2016lzk} (see also~\cite{Harribey:2022esw,Benedetti:2020seh, Gurau:2019qag, Klebanov:2018fzb, Delporte:2020rce} for reviews and references therein) \footnote{It should be noted that, similar to our Eq.~\eqref{eq.sdeq_LO}, such analytic solutions always start by solving self-consistently a large-$N$ Schwinger-Dyson equation for the two point function.}.

We stress that the behavior we encounter here is of a very different nature. The infrared regime we identify does not correspond to a renormalization group fixed point: although the (classically marginal) tetrahedral coupling flows to a fixed value, the renormalized mass is nonzero and larger than a threshold value.

\section{Real tetrahedral coupling} 
\begin{figure}[t!]
\includegraphics[width=0.99\columnwidth]{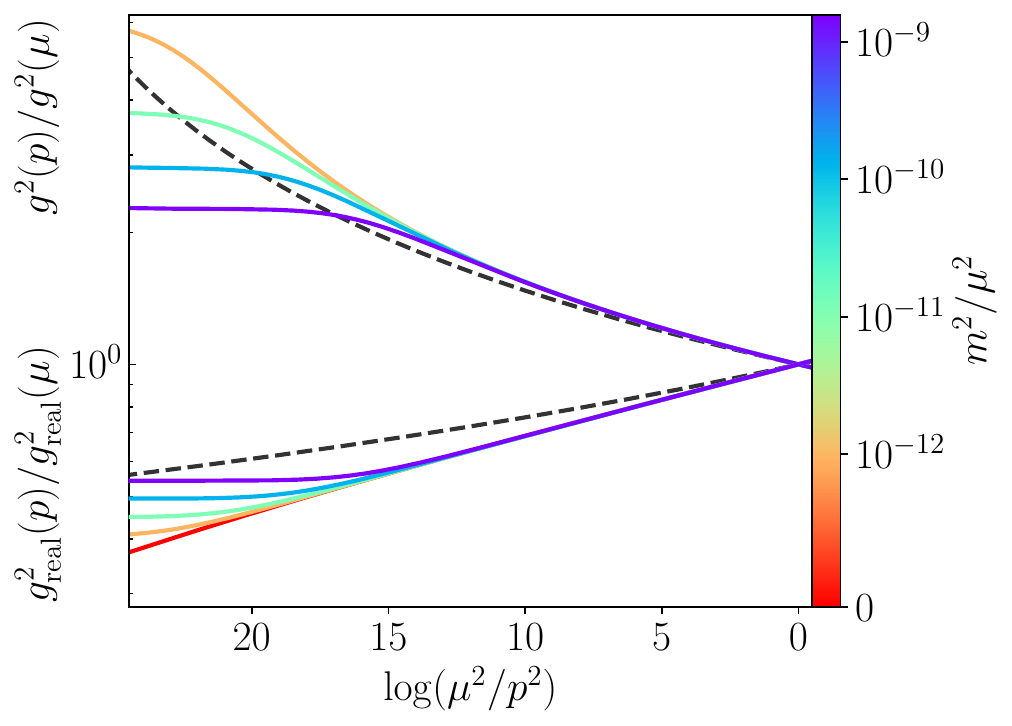}\caption{Comparison of real vs.~imaginary tetrahedral coupling for different renormalized masses and $|g(\mu)|=|g_{\mathrm{real}}(\mu)|$. The corresponding perturbative predictions are shown as dashed black lines. For real tetrahedral coupling also the solution at vanishing renormalized mass is shown (red curve).
}
\label{fig.real}
\end{figure}

If one considers the $O(N)^3$ model with a real tetrahedral coupling in exactly four dimensions, the melonic fixed point of order $\sqrt{\epsilon}$ coincides with the trivial Gaussian fixed point. The perturbative computation, Eq.~\eqref{eq:pertflow} with the sign in the denominator flipped, shows that the tetrahedral coupling vanishes in the IR like $g^2(p)\sim \log\left(\mu^2/p^2\right)^{-1}$ and displays a UV Landau pole at a finite scale. It is not known whether this flow is completed by some nontrivial UV fixed point. In Fig.~\ref{fig.real} we contrast the running in the real and imaginary case obtained by solving Eq.~\eqref{eq.sdeq_LO} with $g^2 \to -g^2_{\mathrm{real}}$. The corresponding two-point functions are discussed in the Appendix. For a real tetrahedral coupling we are able to obtain a self-consistent solution for the propagator with vanishing renormalized mass. This solution exhibits a vanishing coupling for $p\to 0$ (red curve in Fig.~\ref{fig.real}). The self-consistent running decreases faster toward the IR than predicted by perturbation theory (dashed). Similar to the imaginary tetrahedral coupling case, the self-consistent solution with real tetrahedral coupling is sensitive to the presence of a nonvanishing renormalized mass, whereas the perturbative solution is not.

\section{Discussion and outlook} 
Our results show that for an asymptotically free massive scalar field theory in four dimensions quantum fluctuations can generate 
a strong coupling that remains well defined in the infrared. The phenomenon is captured in closed form by the equation for the scalar field two-point correlator in the large-$N$ limit. The wave function renormalization determines the growth of the running coupling toward the infrared. In contrast to the standard perturbative behavior, as long as the mass is above a threshold value, the full nonperturbative coupling remains finite in the infrared. By varying the mass over several orders of magnitude, we find that the infrared value of the coupling grows as the mass is decreased, and exceeds the perturbative estimate.

It is very interesting to consider these results in view of other asymptotically free theories such as QCD, where the nonperturbative generation of a strong-interaction scale by quantum fluctuations is known to have striking phenomenological consequences such as confinement~\cite{Wilson:1974sk}. The gluons in QCD are massless, which is protected by local gauge symmetry, and only gauge-invariant quantities are observable. In particular, in QCD the perturbative notion of a (gauge-variant) coupling ceases to hold in the infrared, where it diverges at the confinement scale $\Lambda_\mathrm{QCD}$. Comparing this to the asymptotically free scalar field theory, the role of $\Lambda_\mathrm{QCD}$ is played by the scale $\mu^*_{\mathrm{pert}}$ in the perturbative tensor model. However, 
the scalar field theory allows one to vary the mass scale and investigate the dynamical build-up of a strong-interaction as the correlation length increases in a regime where the nonperturbative coupling remains well defined and its infrared value determines the physical interaction strength.    

\begin{acknowledgments}
We thank 
J.~Horak,
I.~Klebanov,
T.~Krajewski,
T.~Muller,
J.~Pawlowski,
F.~Popov,
P.~Romatschke,
F.~Sattler,
A.~Tanasa,
G.~Tarnopolsky,
R.~Venugopalan,
and
J.~Wessely
for useful discussions and comments on the draft. The authors acknowledge support by the state of Baden-Württemberg through bwHPC and the German Research Foundation (DFG) through grant no INST 40/575-1 FUGG (JUSTUS 2 cluster), the Heidelberg STRUCTURES Excellence Cluster under Germany's Excellence Strategy EXC2181/1-390900948, the DFG under the Collaborative Research Center SFB 1225 ISOQUANT (Project-ID 273811115), and the European Research Council (ERC) under the European Union’s Horizon 2020 research and innovation program (grant agreement No818066). T.P. acknowledges partial support from the Simons Foundation under Award number 994318 (Simons Collaboration on Confinement and QCD Strings) and thanks Stony Brook 
University and Brookhaven National Laboratory for their hospitality.

\end{acknowledgments}

\begin{appendix}
\begin{widetext}
\renewcommand{\theequation}{A\arabic{equation}}
\renewcommand{\thefigure}{A\arabic{figure}}
\setcounter{equation}{0}
\setcounter{figure}{0}
\makeatletter

\section*{Appendix}

\subsection{Renormalization and two-loop perturbation theory}

The renormalized action of the model writes in terms of the renormalized field $\varphi_{\mathbf{a}}$ as
\begin{align}
 S[\varphi] =  \int d^4 x \bigg\{ & \frac{\tilde Z }{2}  \varphi_{\mathbf{a}}(x)(-  \partial^2 ) \varphi_{\mathbf{a}}(x) + \frac{1}{2} ( m^2 + \delta m^2 ) \varphi_{\mathbf{a}}(x) \varphi_{\mathbf{a}}(x)  \nonumber \\
  &  +\frac{1}{4}\left( ( g_1 +\delta g_1) \hat{P}^{(1)}_{\mathbf{a} \mathbf{b};\mathbf{c} \mathbf{d} } + ( g_2 +\delta g_2)  \hat{P}^{(2)}_{\mathbf{a} \mathbf{b};\mathbf{c} \mathbf{d} }+i (  g +\delta g ) \hat{\delta}^t_{\mathbf{a} \mathbf{b}\mathbf{c} \mathbf{d} }\right) \;\varphi_{\mathbf{a}}(x) \varphi_{\mathbf{b}}(x) \varphi_{\mathbf{c}}(x) \varphi_{\mathbf{d}}(x) \bigg\}\; ,
\end{align}
where the wave function renormalization constant is $\tilde{Z}=1+\delta\tilde{Z}$. The bare and renormalized quantities are related by $\bar\varphi_{\textbf{a}}= \tilde{Z}^{1/2} \varphi_\textbf{a}$, $\bar{m}^2 = \tilde{Z}^{-1}( m^2+\delta m^2 ) $, $\bar{g}  = \tilde{Z}^{-2} (g  +\delta g)$, $ \bar{g}_{1,2}  =\tilde{Z}^{-2}(  g_{1,2}  +\delta g_{1,2})$ and the counterterms $\delta\tilde{Z}, \delta m^2$, $\delta g$ and $\delta g_{1,2}$ ensure that the renormalized correlations are free of divergences. The counterterms are fixed by the renormalization conditions
\begin{equation}
\begin{gathered}
    G^{-1}(0) = m^2\;,\quad Z(\mu)= \frac{G^{-1}(\mu)-G^{-1}(0)}{\mu^2} =1\;, \qquad 
\left.\Gamma^{(4,t)}(p_1,p_2,p_3,p_4)\right|_{p_i^2=\mu^2}=\delta\left(\sum_{i=1}^4 p_i\right)\; g(\mu)\; ,
\end{gathered}
\end{equation}
where $G_{\mathbf{a} \mathbf{b} }(x,y) = \langle \varphi_{\mathbf{a}}(x) \varphi_{\mathbf{b}}(y) \rangle = G(x,y) \delta_{\mathbf{a}\mathbf{b} } $ is the renormalized two-point function and $\Gamma^{(4,t)}$ is the tetrahedral channel of the renormalized four-point function.
The four-point functions in the $\hat{P}^{(1)}$ and $\hat{P}^{(2)}$ channel are fixed similarly.

The function $Z (\mu)$ arising in the renormalization condition above is related to $\tilde Z=\tilde{Z}(\mu)$, the renormalization constant in the renormalized action.
The renormalized propagator is $G^{-1}(p) = \tilde {Z}(\mu) G^{-1}_b(p)$ with $ G^{-1}_b$ the resummed propagator computed in the bare theory, hence $Z(p) = \tilde{Z}(\mu) f_b(p)$ for $f_b$ some function which depends parametrically on the bare parameters. This function is of course divergent, that is it exhibits $1/\epsilon$ poles in $d=4-\epsilon$, or logarithmic divergences with the ultraviolet momentum cutoff $\Lambda$ at $d=4$. Fixing $Z(\mu)=1$ yields
$\tilde Z(\mu) = 1 / f_b(\mu)$ which in turn implies $Z(p) = \tilde Z(\mu) / \tilde Z(p)$. The renormalized two-point function at large-$N$ respects the Schwinger-Dyson equation
\begin{align}\label{eq.sde_appendix}
   G^{-1}(p) =\tilde{Z}p^2+ m^2+\delta m^2+ (g_2+\delta g_2)\int \frac{d^4 q}{(2\pi)^4}\, G(q)  +(g + \delta g)^2 \int \frac{d^4 q}{(2\pi)^4}\frac{d^4 k}{(2\pi)^4}\,  G(q) G(k) G(p+q+k)\;.
\end{align}
Imposing the renormalization conditions and taking into account that in the large $N$ limit $\delta g =0$, see~\cite{Benedetti:2019eyl}, we obtain Eq.~\eqref{eq.sdeq_LO} of the main text.

Keeping the renormalization scale $\mu$ fixed, we are interested in the momentum dependence of the physical couplings of the theory, that is the local parts of the effective action at a symmetric point
\begin{equation}
\left.\Gamma^{(n)}(p_i)\right|_{ p_i^2=p^2}= \delta\left(\sum_i p_i\right) Z(p)^{n/2} g^{(n)} (p) \; . 
\end{equation}
As $\delta g=0$ at large-$N$, the momentum dependent tetrahedral coupling is entirely driven by the wave function, $Z^2(p) g(p) = g(\mu)$. The renormalized propagator $G(p)$ is determined self-consistently by Eq.~\eqref{eq.sdeq_LO}, which in turn fixes the wave function renormalization $Z(p)$ and the running coupling $g(p)$. The running of $g_1$ and $g_2$ is dictated by the running of $g$ and can be extracted from their corresponding Bethe-Salpeter equations, which we will address in future work.

\paragraph{Two loops.} The renormalization group flow  at two-loops is obtained using dimensional regularization and minimal subtraction in $d= 4-\epsilon$ and setting $\epsilon=0$. At two loops and after mass renormalization, Eq.~\eqref{eq.sde_appendix} reads (where we denote the renormalization scale with $s$ here for convenience)
\begin{equation}
G_{\mathrm{pert}}^{-1}(p) = \tilde{Z} p^2+m^2+g^2 s^{2\epsilon}\int \frac{d^d q}{(2\pi)^d}\frac{d^d k}{(2\pi)^d}\,  \frac{1}{(q^2+m^2)(k^2+m^2)}
 \left(\frac{1}{(p+q+k)^2+m^2} - \frac{1}{(q+k)^2+m^2}\right) \;,
\end{equation}
and the sunset (melon) diagram evaluates in a Laurent series in $\epsilon$ \cite{kleinert2001critical}
\begin{equation}
\label{ap.melon_pert}
\begin{aligned}
g^2 s^{2\epsilon}& \int \frac{d^d q}{(2\pi)^d}\frac{d^d k}{(2\pi)^d}\,  \frac{1}{(q^2+m^2)(k^2+m^2)((p+q+k)^2+m^2)}
\\ =&
- \frac{g^2m^2}{(4\pi)^4} \left\lbrace\frac{6}{\epsilon^2} + \frac{6}{\epsilon}\left[\frac32 -\gamma +\log\left(\frac{4\pi s^2}{m^2}\right)\right]  + \frac{p^2}{2m^2\epsilon}+ O(\epsilon^0)\right\rbrace \; .
\end{aligned}
\end{equation}
Up to order $1/\epsilon$ the full propagator at two loops is 
$G_{\mathrm{pert}}^{-1}(p) = \tilde{Z}p^2+m^2 - p^2\frac{g^2}{(4\pi)^4 2\epsilon}$. In the minimal subtraction scheme, renormalization is performed by requiring that at the renormalization scale $s$ both the tetrahedral counterterm $\delta g$  and the wave function counterterm $\delta \tilde Z$ are pure divergences. As $\delta g=0$ we have
\begin{equation}          
g(s) = g\; , \qquad \tilde{Z}(s) = 1 + \frac{g^2}{(4\pi)^4 2\epsilon} \; ,
\end{equation}
where the first equation signifies that the coupling constant $g$ in the renormalized action is exactly the physical four point function (in the tetrahedral channel) at the renormalization  scale $s$.

We note that the minimal subtraction prescription differs from  
imposing the renormalization conditions in Eq.~\eqref{eq.renormCond_prop} by finite terms, but the beta functions up to two loops are prescription independent \cite{zinn2002quantum,Weinberg:1996kr}.
The bare tetrahedral coupling writes $ \bar{g} = s^{\epsilon}g(s) \tilde{Z}^{-2}(s) $. Taking the $s$ derivative at fixed $\bar g$, at two loops we obtain
\begin{equation}
\beta(g) = s \partial_s g(s) = -\epsilon g(s) - \frac{2 g^3(s)}{(4\pi)^4} \;, 
\qquad
\eta= s \partial_s \log(\tilde{Z}(s)) = \frac{\beta(g) \partial_g \tilde{Z}(s)}{\tilde{Z}(s)}  = -\frac{g^2(s)}{(4\pi)^4 } \; .
\end{equation}
We emphasize that we obtained a scale-dependent coupling despite $\delta g=0$, as the flow is driven by the wave function renormalization $\tilde{Z}(s)$, which is nontrivial already at leading order in the large-$N$ expansion. This feature sets melonic tensor field theories, such as $O(N)^3$ at large-$N$, apart from the more standard $O(N)$ vector models at large-$N$. We set $\epsilon =0$ and integrate the flow down from some reference scale $\mu$ in the UV to find
\begin{align}
\label{eq.gZpert}
    g^2_{\mathrm{pert}}(s) = \frac{g^2(\mu)}{1 +  \frac{2 g^2(\mu)}{(4\pi)^4} \log\left(\frac{s^2}{\mu^2}\right)}  \;,
    \qquad
    \tilde{Z}_{\mathrm{pert}}(s) = 
    e^{ -\int_{g(s)}^{g(\mu)} \frac{\eta(g') dg'}{\beta(g')} } \tilde{Z}(\mu)
    =  \left[
    1 +  \frac{2 g^2(\mu)}{(4\pi)^4} \log\left(\frac{s^2}{\mu^2}\right)
    \right]^{-\frac14}  \tilde{Z}(\mu)    \;.
\end{align}
Taking into account the relation between $Z(p)$ and $\tilde Z(p)$, we get
\begin{equation}
 \label{eq.Zpert}
 Z_{\mathrm{pert}}(p) = \left[
    1 +  \frac{2 g^2(\mu)}{(4\pi)^4} \log\left(\frac{p^2}{\mu^2}\right)
    \right]^{\frac14} \;.
\end{equation}

\subsection{Full propagator}
\begin{figure}[t!]
\includegraphics[width=0.46\columnwidth]{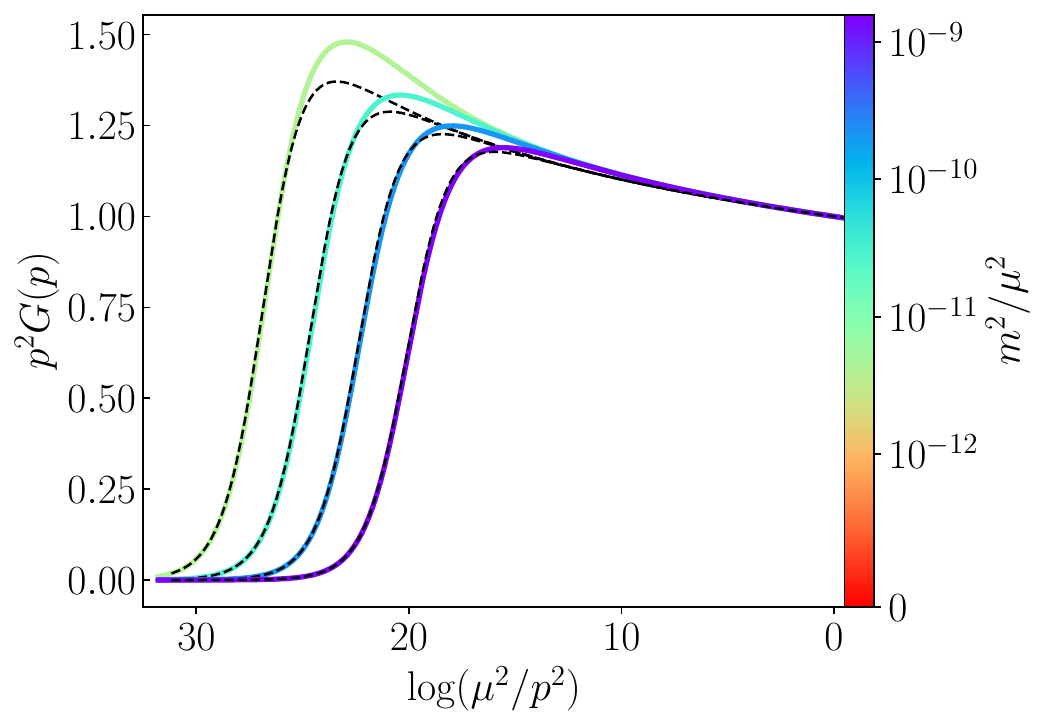}\includegraphics[width=0.45\columnwidth]{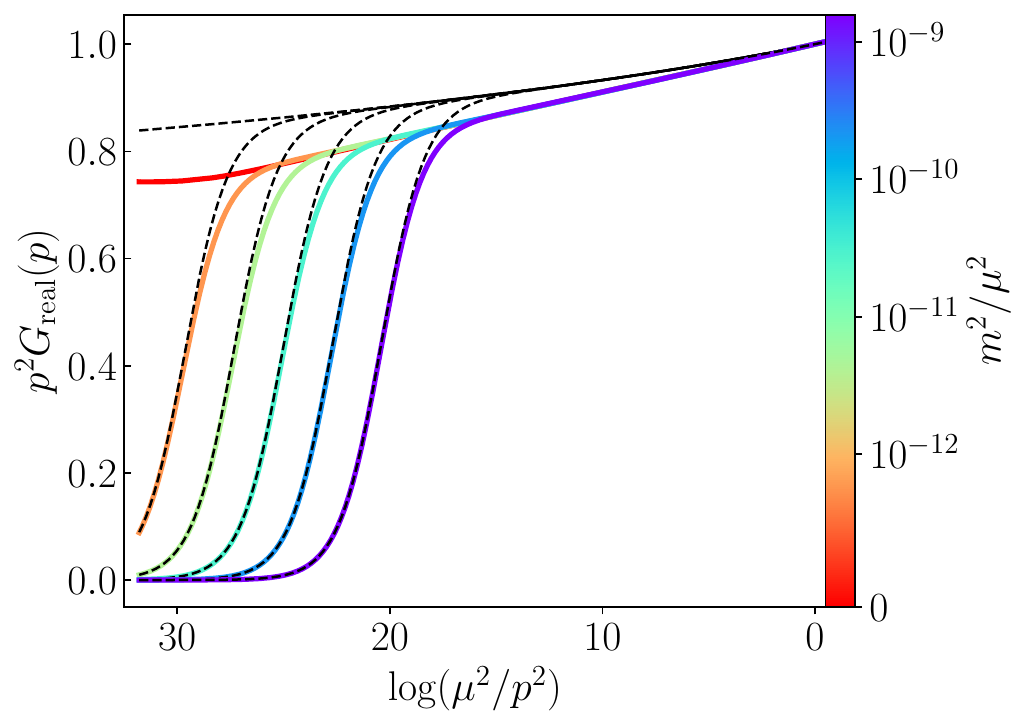}\caption{Propagator with $1/p^2$ momentum dependence scaled out respectively for the case of imaginary tetrahedral coupling (left) and real tetrahedral coupling (right) displayed for various renormalized masses (color scale). The two loop results correspond to the black dashed lines.}
\label{fig.p2G}
\end{figure}
In Fig.~\ref{fig.p2G} we display the renormalized self-consistent propagator for different renormalized masses with the trivial $1/p^2$ momentum dependence scaled out. The flow of couplings discussed in Fig.~\ref{fig.real} in the main text was obtained from the propagators depicted here. 

The case with imaginary tetrahedral coupling is shown on the left. Due to the Gaussian UV fixed point, all the solutions asymptote to $1$ at high momenta. As we decrease $p$ toward the IR, $p^2 G(p)$ increases until the presence of a nonvanishing renormalized mass $m^2$ eventually suppresses the momentum dependence and $p^2G(p)\sim p^2/m^2$ drops to zero. If we decrease $m^2$, we extend the range of scales over which $p^2 G(p)$ grows and we observe an increasingly prominent bump at intermediate momentum scales with associated increasing deviations from the two-loop perturbation theory result
\begin{equation}\label{eq.Gpert}
    p^2 G_{\mathrm{pert}}(p) \simeq \frac{p^2}{m^2 + p^2 \left[1 + \frac{2g^2(\mu)}{(4\pi)^4} \log\left( \frac{p^2}{\mu^2}\right) \right]^{\frac{1}{4}}} \; ,
\end{equation}
which is displayed in dashed black. 
Observe that Eq.~\eqref{eq.Gpert} is only valid for momenta $p^2\geq(\mu^*_{\mathrm{pert}})^2$ with $ \mu^*_{\mathrm{pert}} = \mu \, \exp[-(4\pi)^4/(4 g^2(\mu))] $ since the term in square brackets becomes negative for $p\le \mu^\ast_{\mathrm{pert}}$.

The case with real tetrahedral coupling is shown on the right in Fig.~\ref{fig.p2G}. In contrast to the imaginary case, the propagator is more suppressed for low momenta and we were able to obtain a solution for vanishing mass.
In the perturbative two-loop result for the propagator in Eq.~\eqref{eq.Gpert} only the sign in front of $g^2(\mu)$ changes. This makes the perturbative result well defined for all momenta $p\leq\mu^*_{\mathrm{pert}}$.
These differences are not surprising as for real tetrahedral coupling the Gaussian fixed point is IR attractive. In contrast to the imaginary tetrahedral case, for real coupling it is not known whether there exists a nontrivial UV fixed point and consequently the theory might not exist without a UV cutoff.

\subsection{Numerical implementation}

Implementing the renormalization conditions~\eqref{eq.renormCond_prop} amounts to subtracting the sunset integral evaluated at zero external momentum respectively at the renormalization scale $\mu$. The mass and wave function counterterms are
\begin{align}
\label{ap.counterterm_m2}
    \delta m^2 &=-(g_2+\delta g_2) \int \frac{d^4 q}{(2\pi)^4} G(q) -g^2(\mu) \int \frac{d^4 q}{(2\pi)^4} \frac{d^4 k}{(2\pi)^4} G(q) G(k) G(q+k) \;,
\\
\label{ap.counterterm_Z}
    \delta \tilde Z &= - \frac{g^2(\mu)}{\mu^2} \int \frac{d^4 q}{(2\pi)^4} \frac{d^4 k}{(2\pi)^4} G(q) G(k) \left[G(\mu+q+k) - G(q+k)\right] \;,
\end{align}
leading to the renormalized version of the Schwinger-Dyson equation in Eq.~\eqref{eq.sdeq_LO}
\begin{equation}
\label{ap.sde_renormalized_explicit}
    G^{-1}(p) = p^2(1+\delta \tilde Z)+m^2  + g^2(\mu) I(G;p) \;,
\end{equation}
with
\begin{equation}\label{ap.bigintegral}
I(G;p) =\int \frac{d^4 q}{(2\pi)^4} \frac{d^4 k}{(2\pi)^4} G(q) G(k) \left[G(p+q+k) - G(q+k) \right] \;,
\end{equation}
and $\delta \tilde Z = -g^2(\mu)/\mu^2\,I(G;\mu)$.

We discretize by using a logarithmic $p^2$ grid with $7000$ points and IR/UV cutoffs, respectively $\Lambda_{\mathrm{IR}}/\mu= 1.25\times 10^{-7}$ and $\Lambda/\mu=1.25$, which regularize the integral. 


\subsubsection{Algorithm}
We solve the renormalized Schwinger-Dyson equation for the propagator via fixed-point iteration. The algorithm proceeds as follows:
\begin{enumerate}
    \item We initiate the solver by providing the arbitrary renormalization scale $\mu$, chosen for convenience to lie in the UV with $\mu/\Lambda = 0.8$, and the renormalized parameters $m^2$ and $g(\mu)$. 
    \item Due to asymptotic freedom~\cite{Berges:2023rqa}, 
    we can and do choose the initial ansatz for the propagator to coincide with the  classical one $\left[G^{-1}(p)\right]^{i=0} = p^2+m^2 $.
    Here the superscript $i$ denotes the iteration step.
    \item We calculate the integral $I([G]^i;p)$ in Eq.~\eqref{ap.bigintegral}. This requires some interpolation and extrapolation for $p^2$ values that are not elements of the grid (see next paragraph).
    \item We determine $\delta \tilde Z$ from $I([G]^i;\mu)$.
    \item We evaluate the right hand side of Eq.~\eqref{ap.sde_renormalized_explicit} as $[\mathrm{RHS}]^{i+1} = p^2 (1 - g^2(\mu) /\mu^2  I([G]^i,\mu) )  +m^2 + g^2(\mu) I([G]^i,p)$ and set $[G^{-1}]^{i+1}=\alpha [\mathrm{RHS}]^{i+1}+(1-\alpha)[G^{-1}]^{i}$ with mixing parameter $\alpha=0.2$ to improve the convergence of the algorithm.
    \item We repeat steps 3--5 until apparent convergence is achieved. This is quantified by confirming that the grid-point-wise relative deviation of $[G^{-1}]^{i}$ and $[G^{-1}]^{i+1}$ averaged over all grid points is below a predefined threshold, $10^{-7}$ in our case.
\end{enumerate}

\bigskip

We tested the insensitivity of all displayed results by varying the resolution of the grid and the cutoffs over four orders of magnitude.

 \subsubsection{ Integration}
We use hyperspherical coordinates $(r,\theta,\psi,\phi)\in [0,\infty)\times[0,\pi]\times[0,\pi]\times[0,2\pi]$ and denote $z=\cos(\theta)$, $y=\cos(\psi)$, such that the integral measure on $\mathbb{R}^4$ can be written as $d^4p = \frac12 p^2 \sqrt{1-z^2}\, d(p^2) dzdyd\phi$. Due to spherical symmetry all functions only depend on the invariant $p^2$.
We make use of the fact that the sunset integral
\begin{align}
    M(p) &= \int \frac{d^4 q}{(2\pi)^4} \frac{d^4 k}{(2\pi)^4} G(q) G(k) G(p+q+k) \;,
\end{align}
can be written as two nested convolutions. First, we define
\begin{equation}
    F(p) =\int \frac{d^4 k}{(2\pi)^4} G(p+k) G(k) =  \frac{1}{(2\pi)^3} \int_{\Lambda^2_{\mathrm{IR}}}^{\Lambda^2} dk^2 \int_{-1}^1 dz \;k^2 \sqrt{1-z^2} \;   G(p^2+k^2+|p||k|z)G(k^2)\;,
\end{equation}
and, second, we notice that $I(p)=M(p)-M(0)$ can be computed as
\begin{equation}
\begin{aligned}\label{ap.I(p)}
    I(p)= \int \frac{d^4 q}{(2\pi)^4}  \left( F(p+q)-F(q) \right) G(q)
    =\int_{\Lambda^2_{\mathrm{IR}}}^{\Lambda^2} 
    \frac{ dq^2  }{(2\pi)^3} 
    \int_{-1}^1 dz\,q^2 \sqrt{1-z^2} G(q^2) \left[F(p^2+q^2+|p||q|z)-F(q^2)\right] \;.
\end{aligned}
\end{equation}
The angular $z$-integrals are performed via a Gauss-Chebyshev quadrature with $64$ points and for the $q^2$-integrals we use Gauss-Legendre quadrature~\cite{numericalrecipies} with $7000$ points.

Momenta probed in the convolution range from $0$ to $2\Lambda$, such that $G^{-1}(p)$ and $F(p)$ need to be extrapolated.
 For $|p|<\Lambda_{\mathrm{IR}}$ we set both functions to be equal to their values at $\Lambda_{\mathrm{IR}}$. For $\Lambda<|p|<2\Lambda$ we make use of asymptotic freedom and extrapolate with the momentum dependence inferred from perturbation theory~\footnote{The second equation stems from the integral $\int_{\Lambda_{\mathrm{IR}}<|k|<\Lambda} \frac{d^4k}{(p+k)^2k^2}=\frac{(2\pi)^2}{4}\left[1-\Lambda_{\mathrm{IR}}^2/p^2 + \log(\Lambda^2/p^2)\right]$.}
\begin{equation}\label{ap.extrapolation}
    G^{-1}_{\mathrm{pert}}(p) = m^2+p^2 \left[1+ \frac{2 g^2(\mu)}{(4\pi)^4} \log\left(\frac{p^2}{\mu^2}\right)\right]^{\frac14} \;, \qquad
    F_{\mathrm{pert}}(p) = F(\Lambda)\left[1+\log\left(\frac{\Lambda^2}{p^2}\right)\right] \;.
\end{equation}
The validity of these extrapolations is tested by varying the cutoffs and comparing to the numerical result at high and low momenta. The corresponding numerical result for the convolution $F(p)=\int \tfrac{d^4k}{(2\pi)^4} G(p+k)G(k)$ and the sunset integral with subtracted local part \eqref{ap.bigintegral} are shown in Fig.~\ref{fig.Fbenchmark}.

\begin{figure}[t!]
\begin{minipage}{.48\textwidth}
\includegraphics[height=0.6\columnwidth]{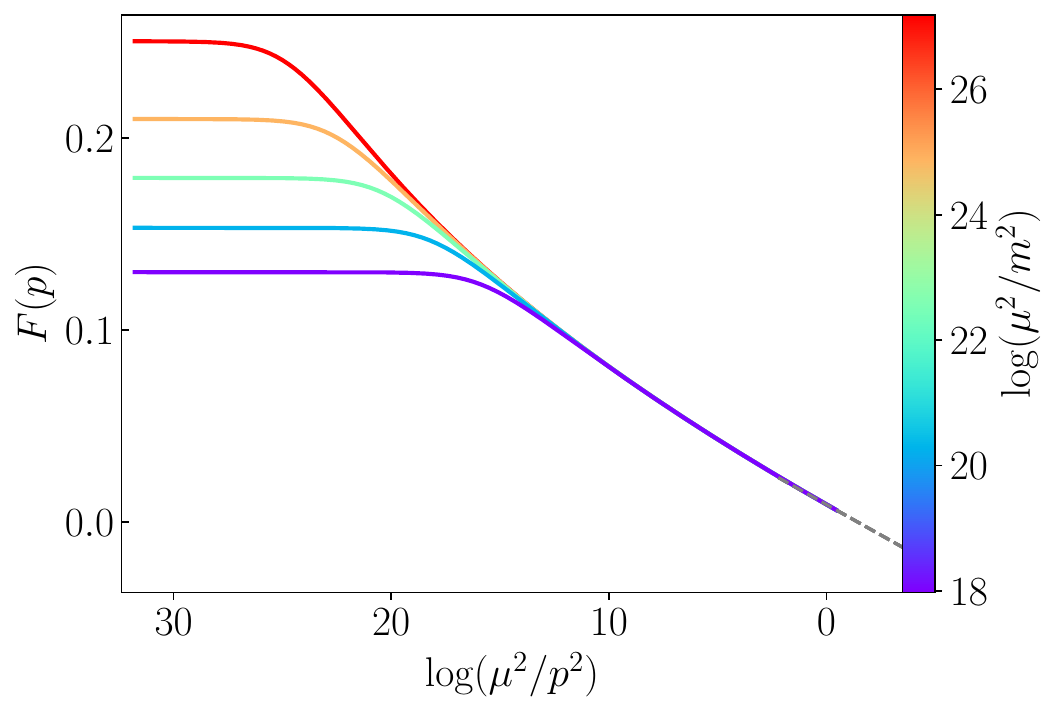}\end{minipage}
\begin{minipage}{0.48\textwidth}
\includegraphics[height=0.6\columnwidth]{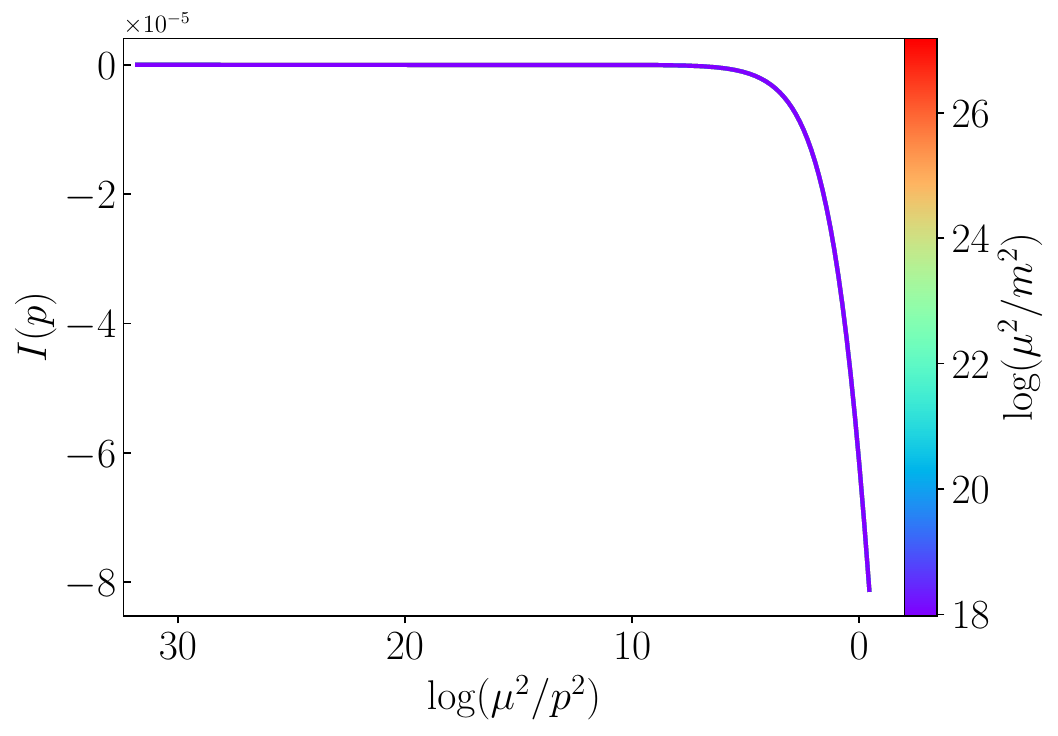}\end{minipage}
\caption{Left: Numerical results for the convolution of two propagators $F(p)=\int \tfrac{d^4 k}{(2\pi)^4} G(p+k)G(k)$ together with the prescribed extrapolation functions \eqref{ap.extrapolation}. The extrapolation for $\Lambda < p \leq 2\Lambda$ is shown in dashed gray. Right: Numerical results for the melon integral with subtracted local part \eqref{ap.bigintegral}. The individual lines for different masses are not distinguishable in this plot.
}
\label{fig.Fbenchmark}
\end{figure}

\end{widetext}
\end{appendix}

\bibliography{master.bib}

\end{document}